\documentclass[conference]{IEEEtran}
\IEEEoverridecommandlockouts
\usepackage{cite}
\usepackage{amsmath,amssymb,amsfonts}
\usepackage{algorithmic}
\usepackage{graphicx}
\usepackage{textcomp}
\usepackage{xcolor}
\usepackage{booktabs}
\usepackage{url}
\usepackage{multirow}
\usepackage{balance}
\usepackage{color, colortbl}

\usepackage{subcaption}
\definecolor{LightGray}{gray}{0.9}

\def\BibTeX{{\rm B\kern-.05em{\sc i\kern-.025em b}\kern-.08em
    T\kern-.1667em\lower.7ex\hbox{E}\kern-.125emX}}
\begin{document}

\newcommand{\ie}{\textit{i}.\textit{e}.,\ }
\newcommand{\eg}{\textit{e}.\textit{g}.,\ }
\newcommand{\cf}{\textit{c}\textit{f}.\ }
\newcommand{\solution}{\textit{DNS Measurement}}
\newcommand\1{\textit{(i)}}
\newcommand\2{\textit{(ii)}}
\newcommand\3{\textit{(iii)}}
\newcommand\4{\textit{(iv)}}
\newcommand\5{\textit{(v)}}
\newcommand\6{\textit{(vi)}}
\newcommand\7{\textit{(vii)}}
\newcommand\8{\textit{(viii)}}
\newcommand\9{\textit{(ix)}}

\newcommand\itemA{\textit{(a)}}
\newcommand\itemB{\textit{(b)}}
\newcommand\itemC{\textit{(c)}}
\newcommand\itemD{\textit{(d)}}
\newcommand\itemE{\textit{(e)}}

%

\title{Traffic Centralization and Digital Sovereignty:  \\ An Analysis Under the Lens of DNS Servers}

\author{\IEEEauthorblockN{Demétrio F. Boeira, Eder J. Scheid, Muriel F. Franco, Luciano Zembruzki, Lisandro Z. Granville}
\IEEEauthorblockA{\\Institute of Informatics (INF),
Federal University of Rio
Grande do Sul (UFRGS),
Porto Alegre, Brazil\\
\{demetrio.boeira, ejscheid, mffranco, lzembruzki, granville\}@inf.ufrgs.br}
}


\maketitle

\begin{abstract}
The Domain Name System (DNS) service is one of the pillars of the Internet. This service allows users to access websites on the Internet through easy-to-remember domain names rather than complex numeric IP addresses. DNS acts as a directory that translates the domain names into a corresponding IP address, allowing communication between computers on different networks. However, the concentration of DNS service providers on the Internet affects user security, privacy, and network accessibility. The reliance on a small number of large DNS providers can lead to (a) risks of data breaches and disruption of service in the event of failures and (b) concerns about the digital sovereignty of countries regarding DNS hosting. In this sense, this work approaches this issue of DNS concentration on the Internet by presenting a solution to measure DNS hosting centralization and digital sovereignty in countries. With the data obtained through these measurements, relevant questions are answered, such as which are the top-10 DNS providers, if there is DNS centralization, and how dependent countries are on such providers.
\end{abstract}



\section{Introduction}
The Internet's Domain Name System (DNS) is a globally hierarchical naming mechanism that enables the association of networks, servers, and services to Internet Protocol (IP) addresses~\cite{mockapetris1988development}. DNS enables, for example, accessing Websites through easy-to-remember domain names rather than IP addresses, meaning that \texttt{wikipedia.org} would be translated to \texttt{208.80.154.224}. The records that map domain names and IP addresses are maintained by authoritative DNS servers that provide authoritative and up-to-date records.

Because deploying a local DNS server requires technical expertise~\cite{dnsServersDeployment}, companies not rarely have been delegating the task of maintaining their authoritative NameServers (NS) records to third-party DNS providers (\eg Cloudflare~\cite{cloudflareDNS} and Akamai~\cite{akamaiDNS}). Such a delegation, which has been increasing over the years~\cite{dnsTrafficCentralization}, led to the current scenario where DNS resolution is concentrated on a small number of large providers. And, for the sake of the business model, each large DNS provider multiplexes its Information Technology (IT) or data center infrastructure among its client companies~\cite{dnsConcentration}. As a result, DNS centralization inevitably leads to security and availability risks, such as user privacy and the inability to resolve domain names in case of an outage or service failure at one of the large providers. The dependency on a few providers creates concerns regarding the \itemA~dependability and \itemB~digital sovereignty of countries, especially considering compliance regulations, such as Europe's General Data Protection Regulation (GDPR) and Brazil's Data Protection Law (LGPD). 

DNS centralization has been widely investigated in the literature. There exist a number of research efforts on assessing the degree of centralization in authoritative DNS servers \cite{dnsTracker} \cite{dnsTrafficCentralization} \cite{doan2021evaluating}, showing \eg that popular domains share the same authoritative DNS servers. Thus, disruptions (\eg due to cyberattacks or sabotage) on DNS infrastructure providers could lead to collateral damages to multiple DNS domains. Although this centralization aspect has been previously addressed, further research on digital sovereignty implications is necessary considering such a DNS dependency. Analyzing digital sovereignty is crucial because it ensures a country's autonomy, control, and security over its digital infrastructure~\cite{digitalsov}. Efforts to quantify the dependency of different countries on DNS providers are, thus, required to uncover possible sovereignty risks for the nations and their critical infrastructures (\eg healthcare, banking, and education sectors), too.

In this paper, we investigate how country code top-level domain (ccTLD) from two conglomerate of countries, \1~Brazil, Russia, India, China, and South Africa (BRICS) and \2~the European Union (EU), are resolved and quantify their dependency on foreign public DNS providers. For that, we define an approach to periodically collect measurements about \texttt{NS} records, \texttt{A} records, and \texttt{AAAA} records in order to find out and map the organizations responsible for managing such providers' infrastructure. These measurements use domains extracted from the Tranco list \cite{trancoList}. Thus, we also analyze how domains are managed and discuss the implications on regulations, compliance, and digital sovereignty. The results show that DNS centralization is a reality and a key concern for digital sovereignty, especially for countries that do not have relevant DNS providers and rely on infrastructure providers from countries or companies with different regulations and~interests.

The rest of this paper is organized as follows. In Section II, we review background knowledge and discuss related work on DNS centralization and digital sovereignty. In Section III, we introduce our \solution{} and its components, including implementation details. In Section IV, we present the evaluation and results, followed by a discussion in Section V. Finally, in Section VI, we close this paper presenting conclusions and discussions about future work.

\section{Background and Related Work}
\label{sec:rw}
Due to the massive damage it may bring to the Internet infrastructure ~\cite{dnsTrafficCentralization}, academia started worrying about and discussing DNS centralization. Some observations found an alarming concentration of DNS traffic, with more than 50\% of the observed traffic being handled by only 10 AS operators~\cite{foremski2019dns}. There is also efforts toward emerging topics to build a responsible Internet \cite{responsible-internet}, which proposes more transparency and trust within networks, independent of vendors and countries that run the underlying infrastructure. Thus, it is clear that companies from the technology and telecommunication sectors have a place to ensure secure communication and a key role in the digital sovereignty.

There are significant concerns about DNS centralization and the impacts it may cause. One big concern is related to performance and how a centralized environment may negatively affect the time-response of DNS in some regions of the globe ~\cite{doan2021evaluating}. Internet Service Providers (ISP) typically operate DNS resolvers for their customers, which means they have access to users' DNS queries and can potentially monitor or manipulate the data, which is definitely a fact to watch. This centralization of DNS resolution can raise privacy and political concerns, mainly if ISPs engage in activities like DNS filtering, censorship, or surveillance~\cite{10091202}.

Furthermore, another two concerns that DNS centralization may raise are security since cyberattacks are evolving and becoming more sophisticated \cite{SecGrid}, including those that target or explore DNS (\eg tunneling, amplification, and flood) to cause technical, economic, and societal impacts \cite{CyberTEA-Paper}. This security concern includes attacks on DNS authoritative servers~\cite{jin2019detection,7208997} and, also, availability of services worldwide, since the phenomenon of centralization is, in addition to being logical, also physical and geographical~\cite{5488355}.

Another concept that can appear from the discussions on DNS centralization is digital sovereignty. Digital sovereignty refers to a nation's ability to control its digital infrastructure, data, and digital technologies within its territorial borders ~\cite{8965469}. It encompasses the idea that countries should be able to shape their digital policies, regulations, and frameworks to protect their national interests, security, and values in the digital realm. Digital sovereignty relies on certain aspects, such as data protection and privacy regulations, domestic digital infrastructure, digital trade and economic policies, and Internet governance~\cite{8756900}. Different works have focused on sovereignty from different perspectives, such as the usage o the decentralization provided by blockchain technology \cite{EderBC} as a potential ally for digital sovereignty \cite{manski2018no}. However, it is unlikely that fundamental changes will become a reality in the short term since, besides enormous technological efforts and associated costs, it depends on convergences between technical and political spheres.

It is important to note that digital sovereignty is a complex and evolving topic, and there are debates around its implementation and potential trade-offs. Striking a balance between digital sovereignty and the benefits of an interconnected global digital ecosystem remains challenging for policymakers worldwide. The question, thus, is how the digital sovereignty of countries is affected by the current Internet and its underlying infrastructure. Thus, it explores DNS and its centralization on few companies and governments to shed light on discussions about digital sovereignty under the lens of DNS.
\section{Measurement Approach}
\label{sec:main}

The approach consists of the mapping of lists of popular Internet domains (\eg based on the publicly available rankings) to its authoritative NSes and organizations behind providing such a service. This allows to identify \1 who provides the correct IPs, \2  which organization operates the NS infrastructure, and \3 to which country and regulations the operator is subjected. For that, the approach combines information from domains (\eg \texttt{A}, \texttt{AAAA}, and \texttt{NS} records) and Autonomous System (AS) records provided by Internet registries (\eg LACNIC, RIPE, and ARIN).

An AS is a network of interconnected computing devices that operate under the same policy. It is often managed by a single entity (\eg ISPs or technology organizations) and is identified by an AS Number (ASN). Each AS manages one or more unique IP ranges, for example, \textit{Wikimedia Foundation Inc.} has an ASN 14907 and manages the IP range \texttt{208.80.152.0/22} in the United States of America and \texttt{185.71.138.0/24} in the Netherlands. Thus, it is possible to associate the IP of any NS to an AS and, consequently, to its operator and region.

Therefore, the approach is able to determine the entire flow from the domain name to the organization handling the AS that manages the IP of the associated NS. This allows to understand the different points where centralization and digital sovereignty risks might occur. For example, the owner of an NS can tamper the DNS records, while the AS operator can outage the communication to make the DNS translation unavailable. In both scenarios, a clear DNS-related dependence can be identified on a few players that maintain the underlying infrastructure (\eg those that operate ASes and NSes). This makes the need to analyze such players and centralization a key pillar for discussing digital sovereignty.



\begin{figure*}[ht]
    \centering
    \includegraphics[width=\linewidth]{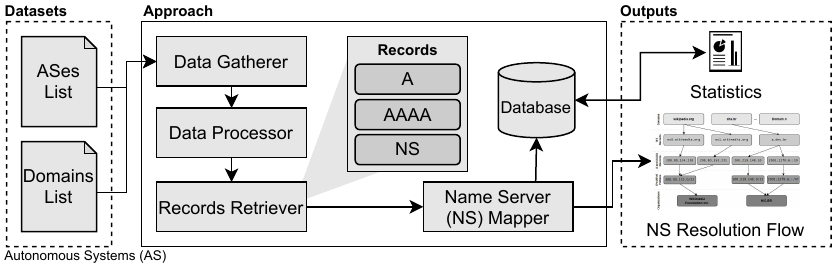}
    \caption{Overview of the Approach to Analysis Domains}
    \label{fig:approach}
\end{figure*}

Figure~\ref{fig:approach} depicts the components that are part of the approach and the flow of information between them. They are organized in three main groups, namely \textbf{Datasets}, \textbf{Approach}, and \textbf{Outputs}. Datasets containing information regarding \textit{Autonomous Systems (AS)} and list of \textit{Domains} are used as inputs for the approach. The ASes responsible for each NS are defined using the list provided by the Center for Applied Internet Data Analysis (CAIDA). For that, it was used the network prefix mapping to AS \cite{CAIDA-IP2AS} and also the mapping of AS to organizations \cite{CAIDA-AS2ORG}. This allows to determine the AS, the organization managing the AS, and, thus, the country/region of DNS providers (based on the IP of the NS). For each measurement, an updated listed of CAIDA is obtained by the \textit{Data Gatherer} and processed to ensure that the correct and most up-to-date information. 

For the domains, the Tranco list \cite{trancoList} is used as dataset since it provides a updated source of top 1 million Websites on Internet based on popularity and access traffic. The list is updated considering a variety of sources, such as Alexa, SimilarWeb and Moz. This offers a reliable and transparent list that can be used to conduct research that needs popular domains. The \textit{Data Gatherer} also obtains the updated Tranco list for each measurement (using a diff approach to identify changes) and the \textit{Data Processor} organize the information of both ASes and domains to be used in further steps.

Next, the \textit{Records Retrieves} analyzes each one of the 1 million domains and retrieves information regarding the A, AAAA, and NS records. For example, for the domain \textbf{wikipedia.org}, the \texttt{A} is \texttt{208.80.154.224}, the \texttt{AAAA} is \texttt{2620:0:861:ED1A::1} and the \texttt{NS} is \texttt{ns0.wikimedia.org}. This information is sent to the mapper to understand the entire path to resolve the DNS in order to build the \textit{NS Resolution Flow} and also collects statistics (\eg organizations concentration, measurement errors, and identified IPs) for further analysis. 

The \textit{NS Mapper} receives the records regarding the domain and obtains the IP of the NS. This information is then used to map the IP to the correspondent AS managing it. For that, the \texttt{A} record can be used in case of an IPv4 prefix or the \texttt{AAAA} for IPv6. Finally, the organization name is obtained by looking at the CAIDA AS organization rank mapping dataset \cite{CAIDA-AS2ORG}, and a complete analysis can be conducted to identify its region and relevant characteristics (\eg regulations and number of ASes being operated). The \textit{NS Mapper} stores information obtained in the \textit{Database} and builds, as output, the \textit{NS Resolution Flow}. This flow shows how the domain is resolved until discovering the organization/company that is managing the infrastructure, which is a point that may directly impact DNS resolutions in case of network disruption. Further, identifying NSes is crucial as they might tamper with DNS records, as they answer the requests in an authoritative manner. 

\begin{figure}[!htpb]
    \centering
    \includegraphics[width=\linewidth]{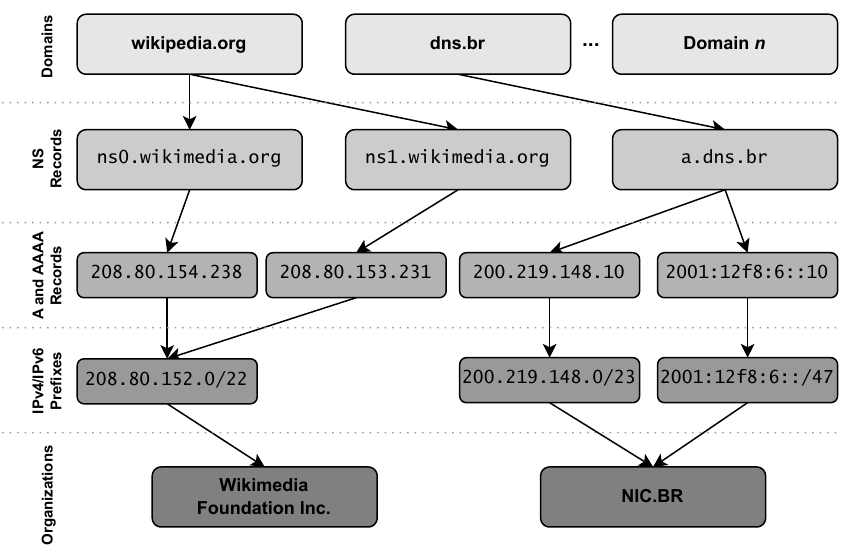}
    \caption{NS Resolution Flow Example}
    \label{fig:resolution-example}
\end{figure}

Figure~\ref{fig:resolution-example} illustrates a graph-like structure of the \textit{NS Resolution Flow} for the domains \textbf{wikipedia.org} and \textbf{dns.br}. In the example, \textbf{wikipedia.org} has two \texttt{NS} records, \1~\texttt{ns0.wikimedia.org} and \2~\texttt{ns1.wikimedia.org}, while \textbf{dns.br} has one \texttt{a.dns.br}. This means these NSes are authoritative servers for these domains and are crucial to their operation. This also applies to the organization that manages the IP addresses and advertise routing information of such servers (\ie their ASes). Such organizations, including the countries they are operating, are retrieved using the \texttt{A} and \texttt{AAAA} records of the NSes by identifying the resolved IPs using their prefixes and mapping them with the AS dataset list. Thus, in the example, \textbf{wikipedia.org} is managed by the \textit{Wikimedia Foundation Inc.}, placed in the United States of America, and \textbf{dns.br} is managed by \textit{NIC.BR}, placed in Brazil.

The approach implementation and results are publicly available at \cite{gitrepository}. Python was used to implement the approach's components, with the \textit{dnspython}~\cite{dnspython}, a Python library to request and manipulate DNS records, being used to implement the \textit{Records Retriever}. The \textit{NS Mapper} connects with a \textit{SQLite3} database to store and manipulate the data required to build the \textit{NS Resolution Flow}. Further, statistics can be retrieved and processed from such a database.
\section{Evaluation and Analysis}
\label{sec:eval}

The measurements considered all the 1~million domains from the Tranco list~\cite{trancoList}, using only the pay-level domains filter,  with the latest list used in the experiments generated on June 16, 2023. To infer the AS names and countries, it was leveraged the CAIDA’s AS-to-organization dataset~\cite{CAIDA-IP2AS}. To conduct the measurements, it was used a six-core AMD Ryzen 5-5500U @ 2.1~GHz with 8~GB of RAM and connected to the Internet using an Ethernet cable to maintain a stable network connection. Its operation system was a Debian 11 ``\textit{bullseye}" stable distribution. 

It is essential to mention that during the experiments, not all domains from the Tranco list were resolved correctly (\eg NS servers not found or incorrectly configured), and their NS or ASN was identified; thus, hindering the possibility of identifying the country where their DNS was managed. However, such limitation does not invalidate the results provided herein.

\subsection{Identifying Top-10 DNS Providers}

Table~\ref{tab:top10-providers} lists the ranking, using 10 positions, of the DNS providers identified during the analysis of centralization aspect of the DNS traffic. The position in the ranking is based on the amount of domains that rely on such DNS provider during the indicated period. Three periods were defined, \textbf{Period 1} from 16/12/2022 to 23/01/2023, \textbf{Period 2} from  23/01/2023 to 13/02/2023, and \textbf{Period 3} from 13/02/2023 to 15/03/2023. As it can be seen in the table, the ranking remained stable during these periods and there was only one change, rows highlighted in gray in the table, in the ranking, where TIGGEE was the 6th during the first two periods but replaced MICROSOFT-CORP-MSN-AS-BLOCK as 7th in the third period.

\begin{table}[ht]
    \centering
    \caption{Top-10 DNS Providers Identified}
    \label{tab:top10-providers}
    \resizebox{\linewidth}{!}{
    \begin{tabular}{cccc}
    \toprule
    \textbf{Position} & \textbf{Period 1} & \textbf{Period 2} & \textbf{Period 3} \\
    \midrule
    1st & CLOUDFLARENET & CLOUDFLARENET & CLOUDFLARENET \\
    2nd &  AMAZON-02 & AMAZON-02 &  AMAZON-02 \\
    3rd &  GODDADY-DNS & GODDADY-DNS & GODDADY-DNS \\
    4th &  ALIBABA-CN-NET & ALIBABA-CN-NET & ALIBABA-CN-NET\\
    5th &  GOOGLE & GOOGLE & GOOGLE \\
    \rowcolor{LightGray}
    6th &  TIGGEE & TIGGEE & MICROSOFT-CORP \\
    \rowcolor{LightGray}
    7th &  MICROSOFT-CORP & MICROSOFT-CORP & TIGGEE \\
    8th &  NSONE & NSONE & NSONE\\
    9th &  IONOS-AS & IONOS-AS & IONOS-AS \\
    10th &  OVH & OVH & OVH \\
    \bottomrule
    \end{tabular}
    }
    \footnotesize \\
    \flushleft
    Gray-highlighted rows indicate a change in the ranking.
\end{table}

Within this context, it was also investigated if the domains of such DNS providers (\eg cloudflare.com) were managed by them or if they relied on services from competitors. Table~\ref{tab:providers-hosting} presents the results of such investigation. The results indicate that not all DNS providers rely on their DNS services for their domains. For example, Amazon, the second largest DNS provider according to Table~\ref{tab:top10-providers}, uses Oracle's DNS services, and Godaddy, which employs its own DNS service but also relies on Akamai's DNS service. However, all the major providers use their own DNS service. This paper does not delve into why \textbf{tiggee.com} is among the largest DNS providers on the Internet but is not listed in the top one million domains according to the Tranco list. Thus, this aspect could be investigated in future~research.

\begin{table}[ht]
    \centering
    \caption{DNS Providers Domains and their Providers}
    \label{tab:providers-hosting}
    \begin{tabular}{ccc}
    \toprule
    \textbf{Domain} & \textbf{DNS Provider} & \textbf{Country} \\
    \midrule
    cloudflare.com &  CLOUDFLARENET & US\\
    amazon.com & ORACLE-BMC-31898 & US\\
    godaddy.com & GODADDY-DNS, AKAMAI-ANS2 & DE, NL\\
    alibaba.com & ALIBABA-CN-NET & US\\
    google.com & GOOGLE & US\\
    tiggee.com & TIGGEE & US\\
    microsoft.com & MICROSOFT-CORP-MSN-AS-BLOCK & US\\
    ns1.com & NSONE & US\\
    ionon.com & IONOS-AS & DE\\
    ovh.com & OVH & FR\\
    \bottomrule
    \end{tabular}
\end{table}

\subsection{Measuring DNS Centralization}

Having identified the top-10 DNS providers that are responsible for hosting the highest amount of domains in the list, one question that arises is if there is an apparent centralization on those providers or if the DNS providing service is highly distributed to avoid Single Point of Failures (SPoF) or monopoly. To answer such a question, it was measured from 16/12/2022 until 15/03/2023, the concentration of domains resolved by the domains listed in Table~\ref{tab:top10-providers}.


\begin{figure}[ht]
    \centering
    \includegraphics[width=\linewidth]{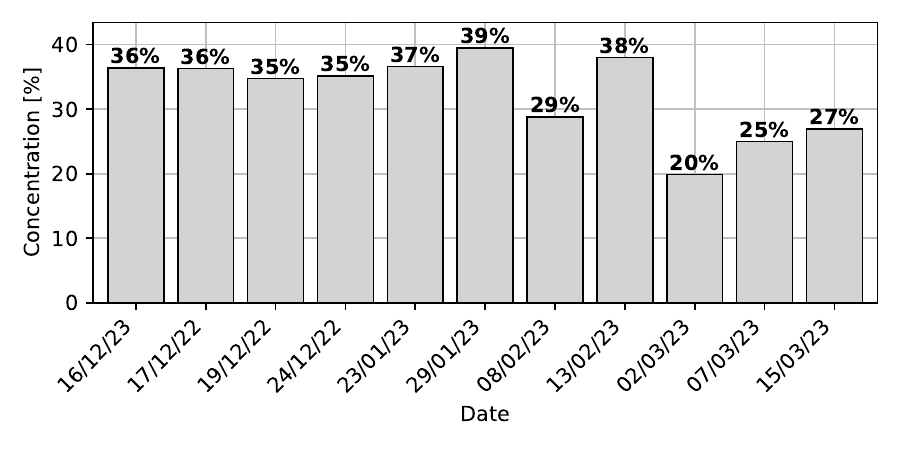}
    \caption{Concentration on Top-10 DNS Providers over 3 Months}
    \label{fig:concentration}
\end{figure}

Figure~\ref{fig:concentration} depicts the results from the performed concentration measurements. In the figure, the \textit{x}-axis represents the date on which the concentration percentage was calculated, and the \textit{y}-axis represents the concentration in the top-10 providers. Considering the period, the average concentration was 30\% of the measured domains. This means that, on average, 30\% of the one million domains of the Tranco list (\ie 300~000 domains) had their DNS records hosted by the top 10 DNS providers (\cf Table~\ref{tab:top10-providers}). Further, considering that such a concentration peaked at 39\% on 29/01/2023 and the fact that it was identified that around 3000 DNS providers were responsible for managing all of the one million domains, there is strong evidence that centralization in the DNS hosting industry is a reality.

\subsection{Analyzing Digital Sovereignty}

Narrowing down the discussion on DNS centralization to a country-based analysis, it is possible to analyze countries' dependency on these providers and quantify how sovereign its Internet infrastructure is in terms of DNS hosting. For that, domains from the Tranco list were selected based on their ccTLD (\eg \texttt{.br} and \texttt{.cn}) and grouped into their political conglomerates.  In total, 91~286 domains from 95~792 domains using the BRICS and EU ccTLDs were resolved, and their DNS hosting organization was identified. This represents 9.1\% and 9.5\% of the Tranco list, respectively. Russia's ccTLD (\texttt{.ru}) represented 59\% of the resolved domains, approximately 54~168 domains. Results from such analysis categorized by these groups are presented in the following~sections.

\begin{figure*}[ht]
\centering
  \begin{subfigure}[b]{.19\linewidth}
    \centering
    \includegraphics[width=.99\textwidth]{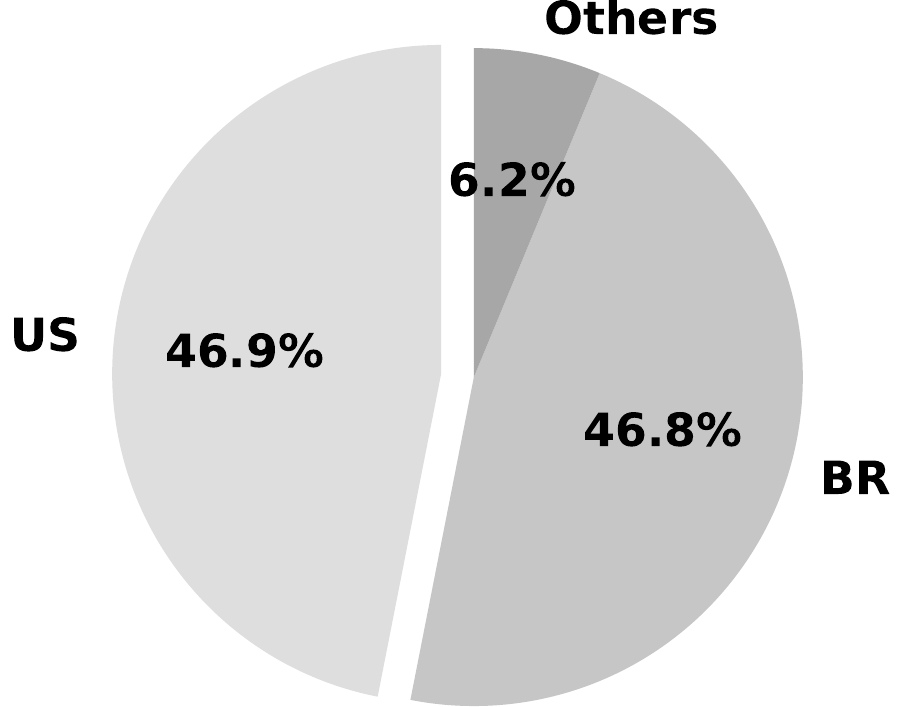}
    \caption{\texttt{.br}}\label{fig:brasil}
  \end{subfigure}%
  \begin{subfigure}[b]{.19\linewidth}
    \centering
    \includegraphics[width=.99\textwidth]{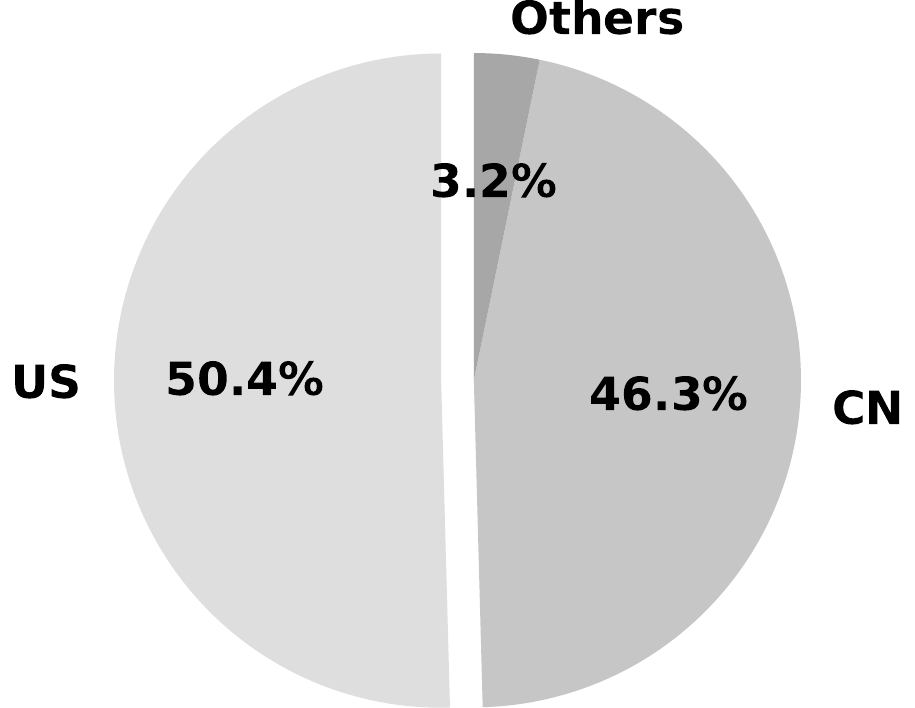}
    \caption{\texttt{.cn}}\label{fig:china}
  \end{subfigure}%
  \begin{subfigure}[b]{.19\linewidth}
    \centering
    \includegraphics[width=.99\textwidth]{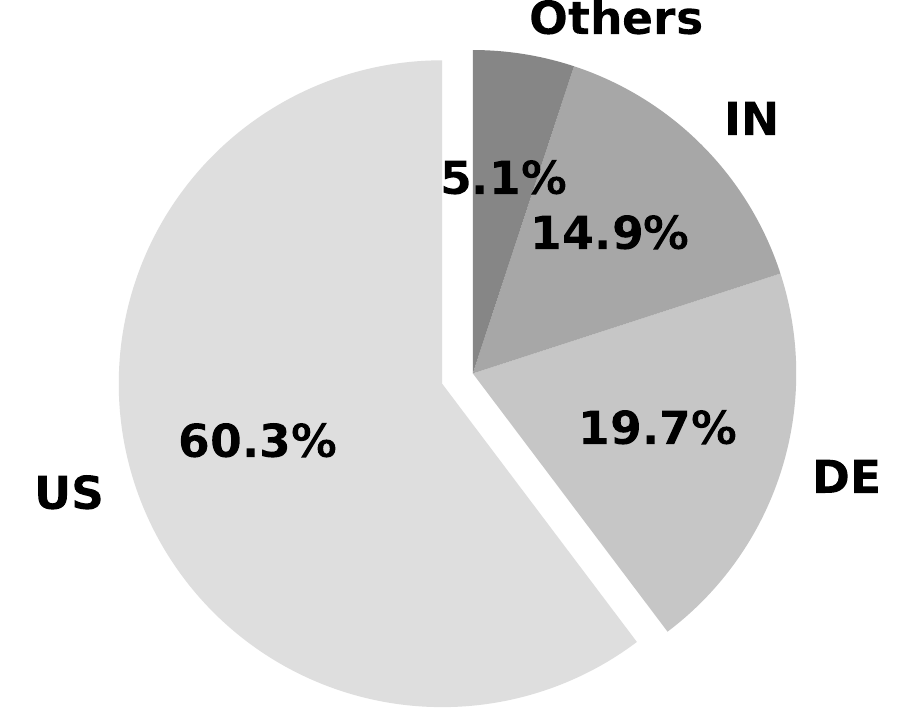}
    \caption{\texttt{.in}}\label{fig:india}
  \end{subfigure} 
    \begin{subfigure}[b]{.19\linewidth}
    \centering
    \includegraphics[width=.99\textwidth]{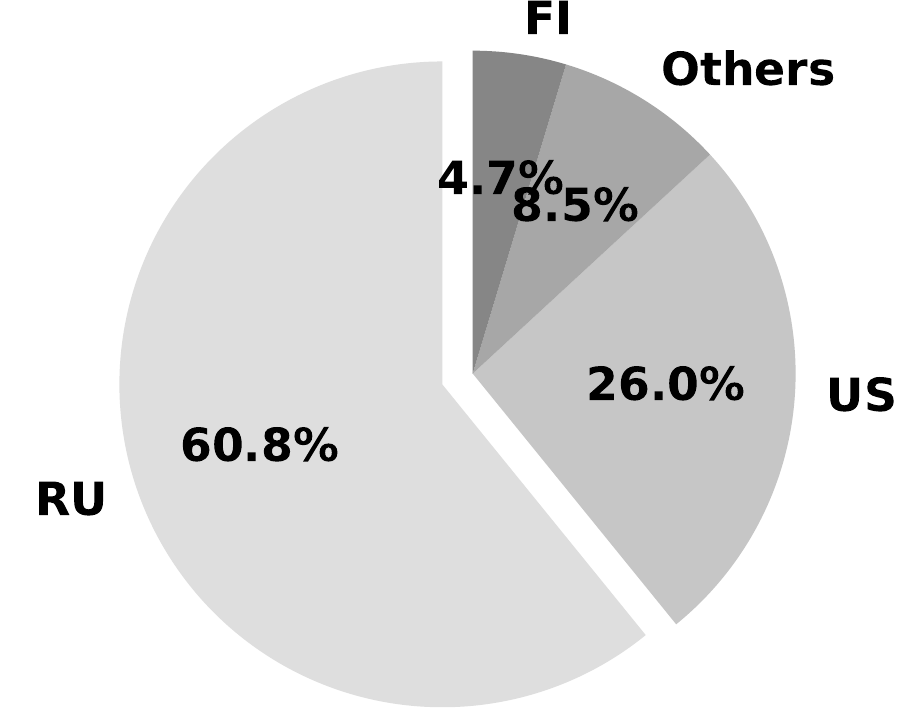}
    \caption{\texttt{.ru}}\label{fig:russia}
  \end{subfigure} 
    \begin{subfigure}[b]{.19\linewidth}
    \centering
    \includegraphics[width=.99\textwidth]{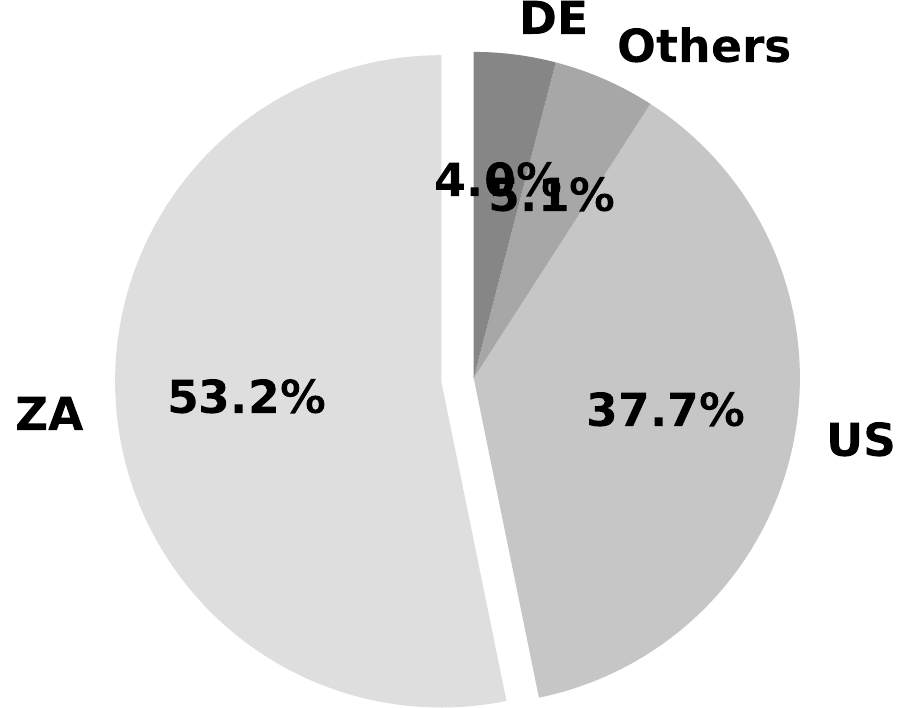}
    \caption{\texttt{.za}}\label{fig:south-africa}
  \end{subfigure} 
  \caption{Results from the BRICS Domains Separated by  ccTLD} \label{fig:brics-hosting}
\end{figure*}

\subsubsection{BRICS Domains}

BRICS represents a conglomerate of five major emerging economies, namely \itemA~Brazil, \itemB~China, \itemC~India, \itemD~Russia, and \itemE~South Africa, formed to promote inter-economic cooperation and inter-political discussions. As BRICS does not have an official ccTLD as Europe, the ccTLD for the BRICS are, respectively, \itemA~\texttt{.br}, \itemB~\texttt{.cn}, \itemC~\texttt{.in}, \itemD~\texttt{.ru}, and \itemE~\texttt{.za}.

Figure~\ref{fig:brics-hosting} depicts the results of the BRICS analysis. Each section of the chart represents the percentage of domains from the defined ccTLD that have their authoritative DNS servers located in the country of the section. The countries are represented as Alpha-2 ISO country codes~\cite{isoCountries}, and countries with less than 4\% of domains were aggregated in the ``Others'' section. For example, in Brazil (\cf Figure~\ref{fig:brasil}), there was a tie between \texttt{.br} domains of the Tranco list that relied on DNS providers from the United States (US) (\ie 46.9\%) and domains that are provided by Brazilian-based companies (\ie 46.8\%), the rest of the share (\ie 6.2\%) were located in other countries (\eg France and Germany).

It is possible to observe that US-based DNS providers, such as Cloudflare, Inc., Amazon.com, Inc., and Google LLC, represent a significant portion of the DNS hosting industry in the BRICS, with India presenting the highest dependence (\ie~60.3\%) of the five nations. The exceptions are Russia (60.8\%) and South Africa (53.2\%), with most domains provided by national DNS companies (\eg Yandex.Cloud LLC for Russia and Xneelo (Pty) Ltd for South Africa). Thus, showing indications of concern regarding digital sovereignty.

\begin{figure}[ht]
    \centering
    \includegraphics[width=0.70\linewidth]{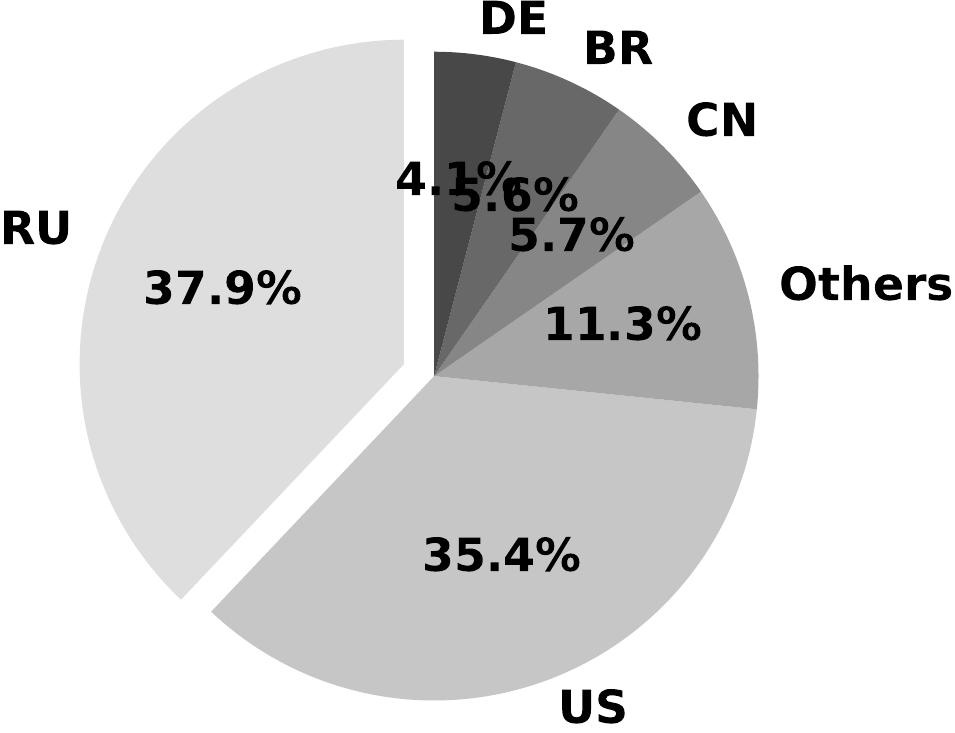}
    \caption{Results from the Aggregated BRICS Domains}
    \label{fig:brics-total}
\end{figure}

Further, to have an overview of the digital sovereignty of the BRICS as a conglomerate, the five countries' results were aggregated and illustrated in Figure~\ref{fig:brics-total}. Russia and the United States appear to host the majority of the domains (\ie a total of 73.3\%), followed by Brazil, China, and Germany. This behaviour is logical considering the division of Figure~\ref{fig:brics-hosting}. Therefore, showing a dystopian view of digital sovereignty, where the BRICS is subject to and dependent on the United States regarding DNS regulations and infrastructure.

\subsubsection{European Union}

\begin{figure}[ht]
    \centering
    \includegraphics[width=0.70\linewidth]{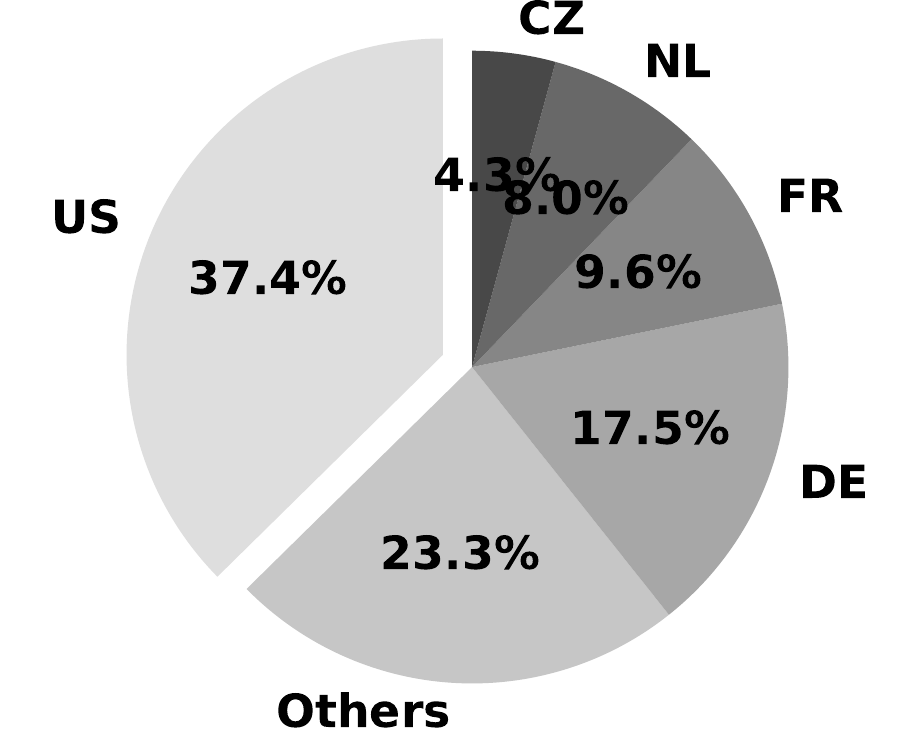}
    \caption{Results from the \texttt{.eu} Domain}
    \label{fig:eu}
\end{figure}

The EU is a political and economic union composed of 27 member states (\eg, Portugal, Spain, France, Italy, Germany, and Hungary) located in Europe. For such countries, the \texttt{.eu} ccTLD was examined. Any person, company or organization within the EU may register domains with this ccTLD. Figure~\ref{fig:eu} illustrates a different scenario than the one from the BRICS (\cf Figure~\ref{fig:brics-total}), where more countries share the DNS hosting infrastructure of the EU. Germany (\ie DE) represents a significant portion given its size and number of DNS hosting providers.

However, the US also concentrates a significant portion of the DNS hosting industry for \texttt{.eu} domains. After Germany, France and the Netherlands appear as major countries hosting DNS domains for Europe, this supports the data presented in Table~\ref{tab:top10-providers}, where OVH, a French cloud computing company, appears as the 10th DNS provider in the ranking. This concentration in a cloud provider might indicate that other services, besides DNS, are being hosted in France and the Netherlands, given the fact that such companies offer more services than DNS, such as virtual machines, Function-on-a-Service (FaaS), and web hosting that require a DNS provider.

\subsection{Hosting Governmental Domains}

One analysis dimension that is highly relevant concerning digital sovereignty and centralization is to investigate where restricted TLDs, such as \texttt{.gov}, are hosted. These domains are intended to be used only by federal government institutions (\eg security agencies and institutes). Thus, their DNS should be hosted within  federal organizations to maintain critical services for citizens and control over the infrastructure during critical periods (\eg global conflicts, pandemics, or sanctions).

\begin{figure}[ht]
    \centering
    \includegraphics[width=1\linewidth,keepaspectratio]{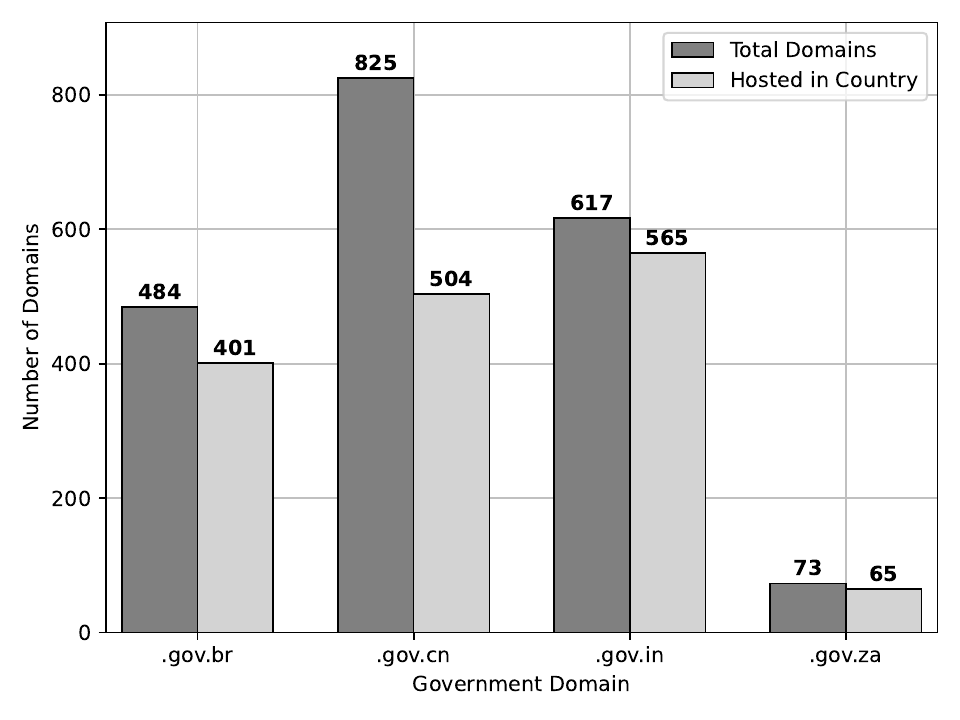}
    \caption{Results from the Analysis of the \texttt{.gov.} Domains}
    \label{fig:gov-brics}
\end{figure}

Figure~\ref{fig:gov-brics} depicts the results from the analysis of the BRICS domains: \texttt{.gov.br}, \texttt{.gov.cn}, \texttt{.gov.in}, and \texttt{.gov.za}. Russia did not present \texttt{.gov} domains in the Tranco list; hence, it is not presented in the results. It can be seen that Brazil's governmental domains are mostly resolved within Brazil, specifically in the Federal Data Processing Service (Serviço Federal de Processamento de Dados - SEPRO, in Portuguese), which is the biggest government-owned corporation of IT services in Brazil. Further, Indian and South African government domains are mostly hosted in their countries, with the National Informatics Centre (NIC) hosting most domains for India and the State Information Technology Agency (SITA) for South Africa. These results show a concern within BRICS about hosting governmental DNS domains for federal services within government organizations to avoid censorship, data leakage and disruption of critical services.

\subsection{Discussion and Key Observations} 

Different insights can be obtained from our experiments under a different lens. From the technical dimension, we have shown that the DNS has a centralization and few players. We also showed that DNS centralization is economic in nature since big techs from developed countries lead the market. Moreover, several economic impacts (\eg business disruption and reputation harm) may happen in companies and governments in case of intentional or non-intentional disruption of DNS underlying infrastructure. Our findings can also be explored from a legal dimension since digital sovereignty involves regulations and actions that can be done by policy-makers based on the technical analysis of the different protocols and dependence (\eg DNS and its centralization on a few companies and countries). The rest of this section provides a discussion on each one of these dimensions.

On the \textbf{technical} dimension, based on the results, it is very straightforward to assume that there is a clear indication of a DNS centralization, which can lead to a scenario where the Internet's infrastructure and management are directly dependent on a few players (\eg governments and companies with different technical and political characteristics). This is not the best scenario, since it can lead to the issues discussed in Section~\ref{sec:rw}, such as security, availability, and performance. Moreover, by allowing such centralization in a given country, region or company, the risk of Internet censorship increases, as such a control can be achieved by injecting fake DNS replies to block access to certain content~\cite{dnsCensorship}. Thus, the DNS infrastructure and its distribution concentrated on a few authoritative servers may lead to Internet outages (due to misconfigurations) and Internet censorship, as the technical enablers for implementing this control are in place.

When discussing the economic dimension of DNS centralization, one point that relates is the possibility of DNS providers profiting from DNS lookup data. \cite{DoHReshaping} advocates that DNS providers do not commercialize such information because of the potential consumer and regulatory backlash of such a monetization. However, suppose the DNS provider's centralization occurs in a country with not-so-well-defined regulations concerning commercializing user-sensitive data. In that case, further monopoly is risky as DNS lookup can be valuable for advertisement. Thus, monitoring and addressing DNS centralization and digital sovereignty is critical to tackling such an economic perspective. Further, most DNS providers (\eg Amazon, Google, and Microsoft) are also major cloud provider companies~\cite{lucianoHostingNOMS22}, where their business is strongly tied to providing a reliable DNS infrastructure to access such cloud instances. However, such a combined service offering leads to a vendor lock-in issue~\cite{vendorLockIn} and even further dependence on their infrastructure, in which companies are subject to such companies' pricing policies.

In addition to these possible economic impacts, DNS centralization also has an economic motivation since big techs (often based in US) offer DNS resolvers and associated services as part of their business core. In 2020, the DNS market was worth USD 372 million, and it is expected to be worth USD 862 million by 2025 \cite{DNSMarket}. This growth expectation is attributed to the growing number of domain name registrations and also Web traffic. Concerns about security, centralization, and digital sovereignty may be part of the marketing and product development strategies for DNS providers and big techs operating the underlying~infrastructure.



Lastly, in the \textbf{legal and political} dimension, there are different efforts from the EU to strengthen the EU's digital sovereignty, such as the GDPR for the idea of data sovereignty and the action plan for more digital sovereignty called by governments of Germany, Estonia, Denmark, and Finland \cite{ActionPlan}. Cybersecurity experts, entrepreneurs, and decision-makers also moved to the discussion to highlight the need to develop and promote digital infrastructures under European technological sovereignty \cite{QuoVadis}. However, even though digital sovereignty is receiving much political attention around the World, the discussions still need to evolve to find a common understanding to succeed. In Brazil, the discussions on the topic are also increasing since discussions on regulations are still needed to increase national cybersecurity and digital sovereignty \cite{BrasilSoberania}. Thus, as seen with these examples and discussions, digital sovereignty is a matter that many stakeholders (\eg governments, companies, and society) have to address from technical, economic, and legal perspectives. Otherwise, digital colonialism may become more prominent and dangerous in the following years, providing mechanisms to increase censorship and digital warfare. 

Thus, as shown in this work and experiments, we advocate that analysis and discussions under different lenses are needed. Besides centralization of protocols such as DNS, cybersecurity, regulations and investments for technology, and mobile communications and its vendors are examples of aspects that must be investigated to lead the discussions of digital sovereignty.

\section{Conclusion and Future Work}
\label{sec:conclusion}

The Domain Name System (DNS) infrastructure plays an essential role in the Internet access infrastructure by allowing content and services to be reached using easy-to-remember names (\ie domains). However, during its development, it was never imagined that such a system would become a market of global proportions. Thus, aspects such as its centralization and governmental regulations were disregarded. In this sense, given its central role in society and concerns regarding the level of control that DNS providers could enforce if the system becomes centralized, understanding and identifying DNS centralization is a key concern.

Thus, in this paper, we presented an approach to measure DNS centralization and digital sovereignty based on DNS domain resolution. The approach relies on a list of 1M popular domains (\ie the Tranco list) and, for each one, identifies the name server responsible for hosting the domain (\ie its authoritative server) and, based on its IP address, maps it to the Autonomous System (AS) managing the IP address. Further, with the AS information, the approach identifies the country in which the AS is located to analyze which regulations the AS is subject to. Consequently, with that information, the approach infers the top-10 DNS providers, the percentage of centralization of the Tranco list in these providers, and also the portion of domains that are managed within their country based on its country-code Top-Level Domain (ccTLD).

Results from the analysis show that most of the top-10 DNS providers identified in the Tranco list are in the US, with Cloudflare being the 1st DNS provider. Further, the analysis of how centralized the DNS hosting industry is revealed that the concentration of domains resolved in the identified top-10 providers peaked at almost 40\%, which shows signals of centralization. Lastly, the results of measuring digital sovereignty in Brazil, Russia, India, China, and South Africa (BRICS) and the European Union (EU) unveiled a scenario where a significant percentage of domains within these countries are not hosted by national companies but hosted on US-based organizations; exceptions being Russia and South Africa. Based on the results, it can be said that not only is DNS centralization occurring on the Internet as previous literature showed (\cf Section~\ref{sec:rw}), but also that countries are becoming less sovereign in terms of control over the national DNS infrastructure.

Considering future work, it is planned to \itemA~analyze such DNS providers distribution with additional countries that are discussing digital sovereignty, \itemB~address the limitations of the work discussed in Section~\ref{sec:eval}, and \itemC~create a tool to analyze DNS providers distribution periodically. Furthermore, our measurement approach can be extended to analyze additional protocols and technologies to provide a more granular technical view of the digital sovereignty landscape.

\bibliographystyle{IEEEtran}
\bibliography{bib/main.bib}
\scriptsize
All links visited on July, 2023

\end{document}